\documentclass[prl,twocolumn,showpacs,10pt]{revtex4}
\usepackage{epsfig}

\begin{document}
\title{Observation of Nonspreading Wave Packets in an Imaginary Potential}
\author{R.~St\"{u}tzle}
\email{stuetzle@kip.uni-heidelberg.de}
\author{M.C.~G\"{o}bel}
\author{Th.~H\"{o}rner}
\author{E.~Kierig}
\author{I.~Mourachko}
\author{M.K.~Oberthaler}
\affiliation{Kirchhoff-Institut f\"{u}r Physik, Universit\"{a}t
Heidelberg, Im Neuenheimer Feld 227, D-69120 Heidelberg, Germany}
\homepage{www.kip.uni-heidelberg.de/matterwaveoptics}
\author{M.A. Efremov,$^1$ M.V. Fedorov,$^1$ V.P. Yakovlev,$^2$
K.A.H. van Leeuwen,$^3$ and W.P. Schleich$^4$}
\affiliation{$^1$General Physics Institute, Russian Academy of
Sciences, 38 Vavilov Street, Moscow, 119991 Russia}
\affiliation{$^2$Moscow Engineering Physics Institute (State
University), 31 Kashirskoe shosse, Moscow, 115409 Russia}
\affiliation{$^3$Eindhoven University of Technology, P.O. Box 513,
5600 MB Eindhoven, The Netherlands} \affiliation{$^4$Abteilung f\"{u}r
Quantenphysik, Universit\"{a}t Ulm, D-89069 Ulm, Germany}

\date{\today}

\begin{abstract}
We propose and experimentally demonstrate a method to prepare a
nonspreading atomic wave packet. Our technique relies on a
spatially modulated absorption constantly chiseling away from an
initially broad de Broglie wave. The resulting contraction is
balanced by dispersion due to Heisenberg's uncertainty principle.
This quantum evolution results in the formation of a nonspreading
wave packet of Gaussian form with a spatially quadratic phase.
Experimentally, we confirm these predictions by observing the
evolution of the momentum distribution. Moreover, by employing
interferometric techniques, we measure the predicted quadratic
phase across the wave packet. Nonspreading wave packets of this
kind also exist in two space dimensions and we can control their
amplitude and phase using optical elements.
\end{abstract}


\maketitle

Nonspreading wave packets have attracted interest since the early
days of quantum mechanics. Already in 1926 Schr\"odinger
\cite{Schrodinger} found that the displaced Gaussian ground state
of a harmonic oscillator experiences conformal evolution because a
classical force prevents the wave packet from spreading. Even in
free space the correlations between position and momentum stored
in an initially Airy-function-shaped wave packet can prevent
spreading \cite{Berry}. Here we propose and experimentally observe
the formation and propagation of nondispersive atomic wave packets
in an imaginary (absorptive) potential accessible in atom optics
\cite{Chudes,Oberthaler,Berry1998}. Although there is no classical
force, there are correlations continuously imposed by Heisenberg's
uncertainty relation resulting in the stabilization of the wave
packet.

Localized wave packets due to stabilization are well known in the
context of periodically driven quantum systems \cite{buchleitner}
and studied with increasing interest for electronic wave packets
in Rydberg atoms \cite{all_theory,maeda,hanson,Chen}. Our approach
to create nondispersive atomic wave packets relies on three
ingredients: (i) an absorption process \cite{note0} cuts away the
unwanted parts of a broad wave creating a packet that is
continuously contracting in position space, (ii) this process
leads due to Heisenberg's uncertainty relation to a broadening in
momentum space and consequently to a faster spreading in real
space, and (iii) the absorptive narrowing and the quantum
spreading are balanced, leading to a nonspreading wave packet. In
the following we will refer to such a wave packet as Michelangelo
packet~\cite{Michelangelo}.

\begin{figure}[h!]
\includegraphics[width=6.51cm]{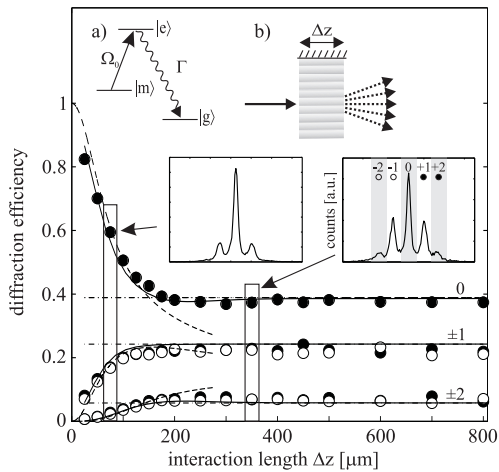}
\caption{\label{fig:1} Formation of a nonspreading Michelangelo
wave packet for the center-of-mass motion of an open two-level
atom (a). The resonant interaction with a standing light wave (b)
leads to an array of harmonic imaginary potentials. The normalized
diffraction efficiencies derived from the momentum distributions
(inset) approach a steady state as a function of the interaction
length $\Delta z$ demonstrating the successful realization of
stationary wave packets. The solid curves result from a numerical
integration of the Schr\"{o}dinger equation
\cite{Fedorov-Efr-Schl-Yak} with the Rabi frequency
$\Omega_0=0.4\Gamma$. The dashed lines correspond to the
Raman-Nath approximation, revealing that the interplay between
absorption and quantum spreading is essential for obtaining  a
steady state.}
\end{figure}

Complex potentials for matter waves \cite{Prentiss_Oberthaler}
emerge from the interaction of near resonant light with an open
two-level system shown in Fig.~\ref{fig:1}(a). For a standing
light wave tuned {\em exactly} on resonance an array of purely
imaginary harmonic potentials arises. When the Rabi
frequency~$\Omega_0$ is of the order of the excited state
linewidth $\Gamma$ the local saturation parameter
$|\Omega_0\sin(kx)/\Gamma|$, and thus the upper level population,
is of the order of unity except in a small vicinity of the field
nodes.
Consequently, our system decays approximately with the rate
$\Gamma$. Therefore, in the time domain $t\gg1/\Gamma$ the atomic
wave function vanishes almost everywhere, except in small vicinity
$\delta x$ of the field nodes. Here the saturation is small and
our open system decays with the rate $(\Omega_0k\delta
x)^2/\Gamma\ll\Gamma$. We estimate the time dependent size $\delta
x$ from the relation $(\Omega_0k\delta x)^2t/\Gamma\sim 1$ and
find $ \delta x(t)\sim (\Gamma/t)^{1/2}/{(k\Omega_0)}.$

The decrease of $\delta x(t)$ is accompanied by an increase of the
width $\delta p(t)\sim 1/\delta x(t)$ in momentum space
\cite{note1} leading to spatial spreading of the wave packet.
Because of competition of the two processes - absorptive
contraction and quantum spreading - the width $\delta x(t)$
reaches its minimal stationary value $\delta x_0$. In this
asymptotic regime the rate $\delta x(t)/t$ of absorptive
contraction is obviously balanced by the rate $\delta p(t)/M$ of
quantum spreading which yields the characteristic time $t_0 \equiv
1/\omega_0 \sim \Omega_0^{-1}
    (\Gamma/\omega_r)^{1/2}$ and the stationary width $\delta
x_0 \sim (M\omega_0)^{-1/2}$ with the recoil frequency $\omega_r
\equiv k^2/(2M)\ll\Gamma$.

The experiments are performed with a slow atomic beam of
metastable argon ($v=50$~m/s) produced with a standard Zeeman
slower. The brilliance of the beam is significantly enhanced with
a 2D-MOT setup \cite{scholz}. The final collimation necessary for
coherent illumination is obtained by two slits ($25~\mu$m and
$10~\mu$m) within a distance of 25~cm. Applying a Stern-Gerlach
magnetic field we select the atoms in the internal state $1s_5$
($J=2$, $m_j=0$). The imaginary potential is realized with a
circularly polarized standing light wave by retroreflecting a
laser beam resonant with the $1s_5$--$2p_8$ transition ($801~$nm).
This setup realizes to a very good approximation an open two-level
system since only 16\% of the excited atoms fall back to the
initial state (in contrast to 32\% without magnetic state
selection). In order to control the interaction length $\Delta z$,
the laser beam passes an adjustable slit. By imaging the slit onto
the retroreflecting mirror we avoid the spoiling effect of light
diffraction. The detection of the metastable argon atoms is
achieved by a microchannel plate detector allowing for spatially
resolved single atom detection utilizing their internal energy
(12~eV). Since the transverse coherence length of the incoming
atomic beam is much larger than the optical wave length, the
outgoing wave function is a coherent array of single Michelangelo
wave packets, resulting in constructive interference in certain
directions. The spatial resolution $\sim 50~\mu$m of our atomic
detector and the free flight distance $\sim 0.5~$m guarantee
clearly resolving the resulting atomic diffraction pattern in the
far field.

The diffraction efficiency is deduced by summing up the detected
number of atoms in angular windows as indicated in the right inset
of Fig.~\ref{fig:1}. After their initial dynamics the wave
packets, i.e., the diffraction efficiencies do not change giving
evidence to the formation of Michelangelo wave packets
\cite{note2}. Our numerical simulations (solid line) of the open
two-level Schr\"odinger equation take into account the
longitudinal as well as the transverse velocity distributions
$\Delta v_l = 10$~m/s and $\Delta v_t =7$~mm/s of the experiment.
For $\Omega_0 = 0.4\Gamma$ we have a very good agreement with our
experimental findings. Since this agreement depends critically on
the Rabi frequency we can determine its absolute value. It is
consistent within a factor of 2 both with a rough estimate, using
the power measurement of the incoming light beam, and with the
overall absorption of the atomic beam.

In order to stress that the interplay between absorptive narrowing
and the quantum spreading is crucial for the formation of the
Michelangelo packet, we have included the result of the Raman-Nath
approximation (dashed lines). Since this approach is only valid as
long as quantum spreading is negligible, it fails to predict the
resulting dynamics after the characteristic time $t_0$.

\begin{figure}[h!]
\includegraphics[width=7.2cm]{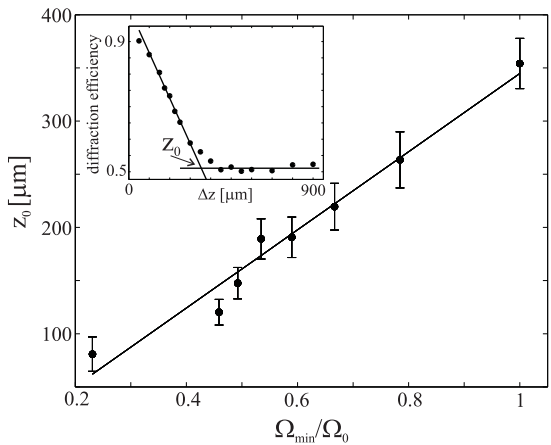}
\caption{\label{fig:2} Experimental verification of the scaling
law $t_0 \sim 1/\Omega_0$ connecting the characteristic time $t_0
\equiv  z_0/v$ when Michelangelo wave packets form and the Rabi
frequency $\Omega_0$. The line is a guide to the eye. We measure
the zeroth order diffraction efficiency as a function of $\Delta
z$ (inset) for different Rabi frequencies. The crossing point
between the linear extrapolation of the short and long-time limits
yields $z_0$.}
\end{figure}

According to the arguments given above, Michelangelo wave packets
emerge after a characteristic time $t_0 \sim 1/\Omega_0$. Our
experimental results shown in Fig.~\ref{fig:2} confirm the
expected scaling with $\Omega_{min}=0.23\Gamma$.

We now show that a Michelangelo wave packet is a complex Gaussian
wave packet with a quadratic phase. For this purpose we recall
\cite{Fedorov-Efr-Schl-Yak} that the solution of the Schr\"odinger
equation
\begin{equation}
 \label{non-ad-main}
 i\frac{\partial}{\partial t}\;\varphi(x,t)=
 \left(-\frac{1}{2M}\frac{\partial^2}{\partial x^2}-
 iU_2(x)\right)\,\varphi(x,t)
\end{equation}
for the metastable state wave function $\varphi(x,t)$ in the
vicinity of $x=0$, where $U_2(x)=M\omega_{0}^2x^2/2$ with
$\omega_0\equiv \Omega_0\,\sqrt{2\omega_r / \Gamma}$
reads~\cite{Berry1998}
\begin{equation}
 \label{result-cordinate}
 \varphi (x,t)=\sqrt{\frac{k/\pi}{\cosh\beta t}}\,
 \exp\left(-\frac{1}{2}\alpha x^2\tanh\beta t\right),
\end{equation}
with $\alpha\equiv M\omega_0 \exp\left(-i\pi/4\right)$ and $\beta
\equiv \omega_0 \exp\left(i\pi/4\right)$.

\noindent Hence, the probability density $|\varphi(x,t)|^2$ is a
Gaussian with the time dependent width $\delta x(t)\equiv \left[
\mathrm{Re}\{ \alpha \tanh (\beta t)\}\right]^{-1/2}$, which for
$\omega_0 t>1$ reaches its minimal stationary value $k\delta x_0
\equiv (\omega_r\Gamma / \Omega_0^2)^{1/4}$.

In this asymptotic regime Eq.~(\ref{result-cordinate}) factorizes
into a product of the time dependent function $\cosh^{-1/2}(\beta
t)$, showing that the Michelangelo probability density decays
exponentially in time with the rate $\Gamma_0 \equiv \omega_0 /
\sqrt{2} \ll \Gamma$, and the position dependent complex Gaussian
$\exp(-\alpha x^2 /2)$ which contains the quadratic phase $ \phi
(x) \equiv M\omega_{0} x^2 / \sqrt{8} $. A Fourier transformation
of this wave packet with the stationary width $\delta x_0$, yields
the asymptotic behavior of the diffraction efficiencies shown in
Fig.~\ref{fig:1} by the dashed-dotted lines and is in perfect
agreement with our experimental findings.

\begin{figure}[h!]
\includegraphics[width=7.0cm]{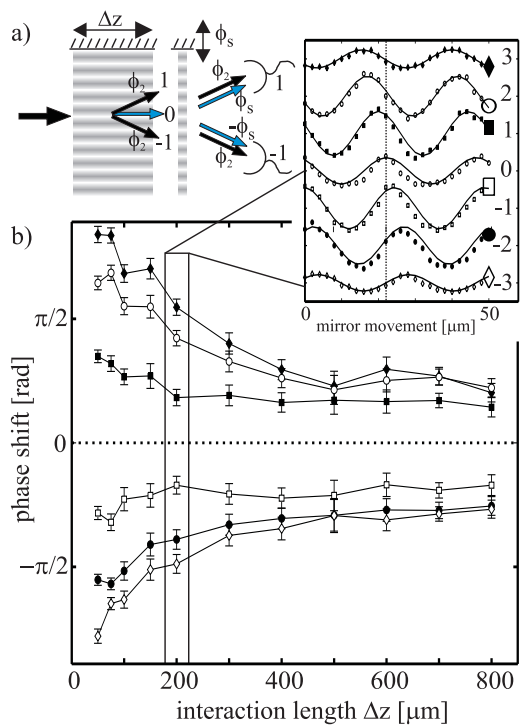}
\caption{\label{fig:3} Measurement of the phase of a Michelangelo
wave packet using an interferometric setup (a) consisting of the
absorptive and probing standing waves. The inset in (b) shows
typical interference patterns for different output directions
obtained by scanning the relative position of the second (thin)
standing light wave for a given interaction length $\Delta z$. For
large values of $\Delta z$ the phase shifts (b) of the different
interferometer outputs relative to the zeroth order level off,
indicating stationary phases of the wave packet.}
\end{figure}

The predicted phase $\phi(x)$ of the Michelangelo packet can be
deduced from the phases of the observed diffraction orders where
the phase of the $n$th order with respect to the zeroth order is
$\phi(n) = -2(\omega_r \Gamma /\Omega_0^2)^{1/2} n^2 \equiv \phi_2
n^2$. To measure the relative phases we realize a compact
interferometer setup shown in Fig.~\ref{fig:3}a. A thin
near-resonant probing standing light wave (waist $30~\mu$m) is
placed directly
behind the array of harmonic imaginary potentials. The wave
function amplitude in each output direction is given as a
superposition of different diffraction orders of the Michelangelo
packet. By changing the relative phase between the two standing
light waves we can measure an interference pattern and thus deduce
the phase evolution as a function of the interaction
length~$\Delta z$.

The interferometric setup employs a probing standing wave at
801~nm realized by beams impinging on the mirror under an angle of
$10^{\circ}$. Thus, moving the mirror allows us to scan the
relative phase $\phi_s$ between the probing and the absorptive
light wave (beating period $25~\mu$m). The presence of a magnetic
field in the interaction region enables us to realize a detuned
(8~MHz) probing wave using the same laser for both standing light
waves but different circular polarizations. By detuning the
probing light wave the total flux through the interferometric
setup is significantly increased in comparison to an exactly
resonant probing field.

In order to deduce the absolute value of $\phi_2$ we evaluate the
interferometer output in the direction of the {\em third}
diffraction order. For our experimental parameters this beam is
always a {\em two-beam} interference of the first and second
diffraction order of the array of Michelangelo packets. In
contrast, the output in lower diffraction order directions is the
result of {\em multiple-beam} interference and does not allow us
easily to deduce the involved phases.

In order to find the phase difference $\phi(2)-\phi(1)$ we have to
eliminate the offset phase arising mainly from the fact that the
probing light field is not infinitely thin. For this purpose we
take the difference between the measured phase in the long-time
limit of the absorptive wave ($\Delta z > 400~\mu$m) and the phase
for the experimentally achievable shortest interaction length
(50~$\mu$m). For the Rabi frequency $\Omega_0 = (0.23\pm
0.02)\Gamma$ we find the experimental value
$|\phi(2)-\phi(1)|=1.70\pm 0.17$, which is in agreement with the
prediction of the numerical integration $\phi(2)-\phi(1)=3\phi_2$,
that is $|\phi_2|=0.57\pm 0.1$. Moreover, the characteristic
length $z_0 \sim 400~\mu$m for leveling off the phases coincides
with the one for leveling off the diffraction efficiencies.
Furthermore, by increasing the Rabi frequency to
$\Omega_0=(0.4\pm0.05)\Gamma$ we experimentally deduce
$|\phi_2|=0.32\pm 0.08$, which is in very good agreement with the
prediction of the numerical integration $|\phi_2|=0.27 \pm 0.04$.

So far we have concentrated on wave packets in $D=1$ spatial
dimensions. A straightforward generalization to $D=2$ relies on
two orthorgonal linear polarized standing waves interacting with
the appropriate atomic transitions and leads to the potential
$-iM(\omega_x^2 x^2+\omega_y^2 y^2)/2$ near the nodes. The
frequencies $\omega_x$ and $\omega_y$ depend on the field
intensities. A nonorthorgonal configuration provides even
additional parameters to control the form of the emerging
two-dimensional Michelangelo wave packet.

We emphasize that Michelangelo wave packets are not restricted to
the Gaussian form, Eq.~(\ref{result-cordinate}), originating from
the quadratic potential $U_2$ in Eq.~(\ref{non-ad-main}). Indeed,
with an
\begin{figure}[t!]
\includegraphics[width=8.4cm]{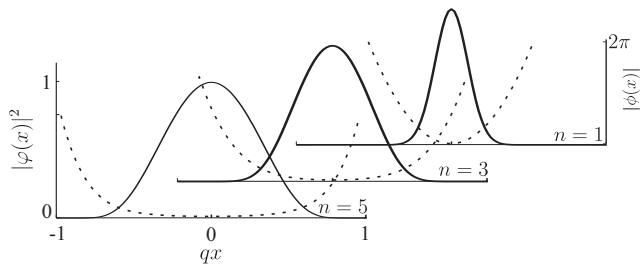}
\caption{\label{fig:4} Probability density $|\varphi(x)|^2$ (solid
line) and absolute value of phase $|\phi(x)|$ (dotted line) of
Michelangelo wave packets in the Potential $U_{2n}$.}
\end{figure}
appropriate mask \cite{mutzel} we can create almost any behavior
of the mode function close to the node, leading, for example, to a
power law potential $U_{2n}(x) = (\Omega_0^2/\Gamma) (qx)^{2n}$.
Here $q\ll k$ determines the characteristic width of $U_{2n}$.

The Michelangelo wave packets shown in Fig.~\ref{fig:4} for
$n$=1,3, and 5 are the "ground" state eigenfunctions of the
corresponding stationary non-Hermitian Hamiltonians and can be
obtained numerically. In the asymptotic regime only these
functions survive because their complex energy "eigenvalues" have
the smallest imaginary parts. Moreover, applying the general
arguments above to the case of $U_{2n}$ yields the following
characteristic time and width:
\begin{equation}
t_0\sim
\frac{1}{\Omega_0^{2/(n+1)}}\left(\frac{\Gamma}{\tilde{\omega}_r^n}\right)^{1/(n+1)}
\;\, \text{and} \quad\, q \delta x_0 \sim
\left(\frac{\Gamma}{\Omega_0^2 t_0} \right ) ^{1/2n},
\end{equation}
where $\tilde{\omega_r}=q^2/(2 M)$. These scaling behaviors have
been confirmed by numerical integration of the corresponding
Schr\"odinger equation. We note that for $n=1$ these expressions
reduce to the ones of the previous case. For $n\rightarrow \infty$
the potential $U_{2n}$ takes on the shape of a box, $t_0$ is
independent of $\Omega_0$, and $\delta x_0$ is solely given
by~$q$.

In conclusion we present a new class of nonspreading wave packets
resulting from the interplay between absorptive narrowing and
quantum spreading. The developed theoretical description explains
the experimental observation of both the phase and the amplitude
of the wave packet quantitatively. The experimental realization of
imaginary potentials strongly relies on spontaneous decay
processes. Nevertheless, we show that coherence is maintained and
can even be employed for deducing the phase of the Michelangelo
packets. Since the wave packet arising in the long-time limit is
weakly dependent on the initial wave function, this process is a
robust tool for generating wave packets with well-defined
amplitude and phase for further experiments.

We acknowledge fruitful discussions with A. Buchleitner and thank
M. St\"{o}rzer for his commitment in the early stage of the experiment
which was funded by Optik-Zentrum Konstanz, Center for Junior
Research Fellows in Konstanz, by Deutsche Forschungsgemeinschaft
(Emmy Noether Program), and by the European Union, Contract No.
HPRN-CT-2000-00125. MVF, WPS, and VPY also thank the Alexander von
Humboldt-Stiftung for its generous support during the course of
this project, especially for the Humboldt-Kolleg at Cuernavaca,
Mexico. This work was partially supported by the Landesstiftung
Baden-W\"{u}rttemberg and the Russian Foundation for Basic Research
(grants no. 02-02-16400, 03-02-06145, 04-02-16734).


\begin{thebibliography}{99}
\bibitem{Schrodinger} E. Schr\"{o}dinger, Naturwissenschaften {\bf
14}, 664 (1926).

\bibitem{Berry} M.V. Berry and N.L. Balazs, Am. J. Phys. {\bf 47},
264 (1979).

\bibitem{Chudes} D.O. Chudesnikov and V.P. Yakovlev, Laser Phys. {\bf 1}, 110 (1991).

\bibitem{Oberthaler} M.K. Oberthaler et al., Phys. Rev. Lett. {\bf 77}, 4980
(1996).

\bibitem{Berry1998} M.V.~Berry and D.H.J.~O'Dell, J. Phys. A {\bf
31}, 2093 (1998).

\bibitem{buchleitner} A.~Buchleitner, D.~Delande, and J.~Zakrzewski, Phys. Rep. {\bf 368}, 409
(2002), and references therein.

\bibitem{all_theory} G.P. Berman and G.M. Zaslavsky, Phys. Lett. A
{\bf 61}, 295 (1977); K. Richter and D. Wintgen, Phys. Rev. Lett.
{\bf 65}, 1965 (1990); J. Henkel and M. Holthaus, Phys. Rev. A
{\bf 45}, 1978 (1992); D. Delande and A. Buchleitner, Adv. At.
Mol. Opt. Phys. {\bf35}, 85 (1994); I. Bialynicki-Birula,
M.~Kalinski, and J.H.~Eberly, Phys. Rev. Lett. {\bf 73}, 1777
(1994); A. Buchleitner and D. Delande, Phys. Rev. Lett. {\bf 75},
1487 (1995); M.V.~Fedorov and S.M.~Fedorov, Opt.~Express {\bf 3},
271 (1998); M. Kalinski et al., Phys. Rev. A {\bf 67}, 032503
(2003).

\bibitem{maeda} H. Maeda and T.F. Gallagher, Phys. Rev. Lett. {\bf 92},
133004 (2004).

\bibitem{hanson} L.G.~Hanson and P.~Lambropoulos, Phys. Rev. Lett. {\bf 74}, 5009
(1995).

\bibitem{Chen} X. Chen and J.A. Yeazell, Phys. Rev. Lett. {\bf 81}, 5772
(1998).

\bibitem{note0} Absorption also plays a crucial role in the
proposal \cite{hanson} for and in the experiment \cite{Chen} with
an electronic nonspreading wave packet in a two-electron atom
using atomic mode locking by loss modulation.

\bibitem{Michelangelo} For Michelangelo, sculpturing means
"releasing the desired form from a block of marble by cutting away
unwanted materials." See, for example, D. Preble,  {\em Artforms}
(Harper $\&$ Row, New York, 1978).

\bibitem{Prentiss_Oberthaler} K.S.~Johnson et al., Science {\bf
280}, 1583 (1998); M.K. Oberthaler et al., Phys. Rev. A {\bf 60},
456 (1999); A.~Turlapov et al., Phys. Rev. A {\bf 68}, 023408
(2003).

\bibitem{note1} Throughout the paper we use $\hbar\equiv 1$.

\bibitem{scholz} A.~Scholz et al., Opt. Comm. {\bf 111}, 155
(1994).

\bibitem{Fedorov-Efr-Schl-Yak} M.A. Efremov et al., Laser Phys. {\bf 13}, 995 (2003);
M.V. Fedorov et al., JETP {\bf 97}, 522 (2003).

\bibitem{note2} This conclusion is only valid since our interferometric experiment
discussed below exclude additional dynamics of the relative phases
between diffraction orders.

\bibitem{mutzel} U. Drodofsky et al., Appl. Phys. B {\bf 65}, 755
(1997); M. M\"{u}tzel et al., Phys. Rev. Lett. {\bf 88}, 083601
(2002).
\end{thebibliography}
\end{document}